\documentclass[12pt,english]{article}
\usepackage[T1]{fontenc}
\usepackage[latin1]{inputenc}
\usepackage{babel}
\usepackage{epsf} 

\makeatletter

\providecommand{\LyX}{L\kern-.1667em\lower.25em\hbox{Y}\kern-.125emX\@}

\usepackage[T1]{fontenc}
\usepackage[latin1]{inputenc}
\usepackage{babel}

\makeatletter

\usepackage[T1]{fontenc}
\usepackage[latin1]{inputenc}
\usepackage{babel}

\makeatletter

\usepackage[T1]{fontenc}
\usepackage[latin1]{inputenc}
\usepackage{babel}



\makeatother

\makeatother

\makeatother
\begin{document}

\title{The Casimir energy of a massive fermionic field confined in a \( d+1 \)
dimensional slab-bag }

\author{E. Elizalde\( ^{\dagger } \)\\
 Institut d'Estudis Espacials de Catalunya (IEEC/CSIC) \\
 Edifici Nexus, Gran Capit\`{a} 2-4, 08034 Barcelona \ \& \\
 Departament d'Estructura i Constituents de la Mat\`{e}ria \\
 Facultat de F\'{\i}sica, Universitat de Barcelona \\
 Av. Diagonal 647, 08028 Barcelona, Spain\\
 and \\
 F. C. Santos\( ^{\ddagger } \) and A. C. Tort\( ^{\flat ,\star } \)\\
 Instituto de F\'{\i}sica - Universidade Federal do Rio de Janeiro
\\
 C.P. 68528, Rio de Janeiro CEP 21945-970 Brasil}

\date{\today{}}

\maketitle
\begin{abstract}
\noindent We evaluate the fermionic Casimir effect associated with
a massive fermion confined within a planar \( d+1 \) dimensional
slab-bag, on which MIT bag model boundary conditions of standard type,
along a single spatial direction, are imposed. A simple and effective
method for adding up the zero-point energy eigenvalues, corresponding
to a quantum field under the influence of arbitrary boundary conditions,
imposed on the field on flat surfaces perpendicular to a chosen spatial
direction, is proposed. Using this procedure, an analytic result is
obtained, from which small and large fermion mass limits, valid for
an arbitrary number of dimensions, are derived. They match some known
results in particular cases. The method can be easily extended to
other configurations. \vskip 0cm
\end{abstract}
\vfill

{\small
\noindent \( ^{\dagger } \) e-mail: elizalde@ieec.fcr.es \\
 \( ^{\ddagger } \) e-mail: filadelf@if.ufrj.br \\
 \( ^{\flat } \) e-mail: tort@if.ufrj.br \\
 \( ^{\star } \) Present address: Institut d'Estudis Espacials de
Catalunya (IEEC/CSIC) Edifici Nexus 201, Gran Capit\`{a} 2-4, 08034
Barcelona, Spain. E-mail address: visit11@ieec.fcr.es  }

\section{Introduction}

The macroscopically observable vacuum energy shift, associated with
a quantum field, is the regularised difference between the vacuum
expectation value of the corresponding Hamiltonian with and without
the external conditions demanded by the particular physical situation
at hand. At the one-loop level, when the external conditions are represented
by boundary conditions, this leads to the usual Casimir effect \cite{Casimir48}
-- see Ref. \cite{BMohideenM2001} for an updated review on the theoretical
and experimental aspects of this remarkable phenomenon.

The physical fact that only bound states of quarks are experimentally
observed, led to the model-dependent idea of the total confinement
of the quark fields, through the so-called bag model. In general,
the bag is modeled by the interior of a spherical shell within which
the quarks and gluon fields are confined. There are several important
calculations concerning the vacuum energy associated with the spherical
bag model \cite{BagModel}, as well as results concerning the influence
of different boundary conditions imposed on the confined fermionic
and gluonic fields \cite{BagModelBC}. Planar bag models with standard
MIT boundary conditions (no fermionic currents through the surface
of the bag) have been also considered in the literature, in particular,
the case of a massless fermionic field was considered in Ref. \cite{Johnson75}
and its extension to \( d+1 \) dimensional was discussed in Ref.
\cite{Svaiteretal1999}. The massive case in three spatial dimensions
was considered in Ref. \cite{MamaevTrunov80}.

In the context of the evaluation of the zero point energies associated
with confined quantum fields, some configurations, which depend on
the nature of the quantum field, the type of spacetime manifold and
its dimensionality, and the specific boundary condition imposed on
the quantum field on certain surfaces, lead to relatively simple spectra,
easy to deal with, but others lead to rather complex ones. The heart
of the matter in these calculations is the (physically meaningful)
evaluation of the spectral sum that results at the one-loop level
from the definition of the Casimir energy. This evaluation requires
regularization and renormalization, and recipes for accomplishing
this task range from the relatively simple cutoff method, employed
by Casimir himself \cite{Casimir48}, to a number of powerful and
elegant generalised zeta function techniques \cite{Elisalde94}. Contour
integral representations of spectral sums are a great improvement
in the techniques of evaluating zero-point energies; they are especially
useful when the spectra are not simple, and have been employed profusely
before \cite{relatedpapers}.

Here, our aim is twofold. First, with the same objectives as in Refs.
\cite{relatedpapers} in mind, we want to generalise to \( n+1 \)
dimensions a simple and effective way of evaluating the one-loop vacuum
energy under external conditions, based on well-known theorems of
complex analysis, namely, the Cauchy integral formula and the Mittag-Leffler
expansion theorem in one of its simplest versions \cite{ST2002,FO2001}.
The method we will present, although very simple, is of sufficient
generality so as to be successfully applied to a variety of cases.
Second, we want to apply our basic formula to the case of the \( n+1 \)
dimensional slab-bag, in which the quark field is constrained into
a region bounded by two infinite hyperslabs separated by a fixed distance
\( \ell  \). MIT boundary conditions (see Ref. \cite{Johnson75}
and references therein) will be imposed on the quantum field. There
are no fermionic field oscillations outside the interior region of
the slab-bag, a situation that must be compared with the one in which
the boundary conditions split the spacetime in several regions, and
the zero-point oscillations of the relevant quantum fields are present
in all of them. This example is one of the simplest one can think
of that leads already to a relatively complex spectrum, and serves
as a convenient test of the calculational tool that we present here.
The paper is divided as follows. In Sect. 2 we introduce the method,
and get the basic formula. In Sect. 3, we apply our main result to
the problem in hand, and obtain its exact solution in terms of Whittaker
functions. Finally, we consider some particular limits and check coincidence
with some other results. The last section is devoted to final remarks.
We employ natural units (\( \hbar =c=1 \)).

\section{The unregularised Casimir energy and a very simple sum formula}

Consider a quantum field living in a \( d+1 \) dimensional ultrastatic
flat spacetime under boundary conditions imposed on two hyperplanes
of area \( L^{d-1} \) kept a distance \( \ell  \) apart. Hence the
motion along one of the spatial directions, say, the \( {\mathcal{OX}}_{d} \)-axis,
is restricted. Suppose that the condition \( L\gg \ell  \) holds.
At the one loop-level, the (unregularised) Casimir energy is given
by \begin{equation}
\label{unregenergy}
E_{0}\left( d\right) =\alpha \left( d\right) \, \frac{L^{d-1}}{2}\int \sum _{n}\frac{d^{d-1}p_{\bot }}{(2\pi )^{d-1}}\Omega _{n},
\end{equation}
 where \( \alpha \left( d\right)  \) is a dimensionless factor that
counts the number of internal degrees of freedom of the quantum field
under consideration, \( p_{\bot }=\sqrt{p_{1}^{2}+p_{2}^{2}+\cdots +p_{d-1}^{2}} \),
and \begin{equation}
\Omega _{n}:=\sqrt{p_{\bot }^{2}+\frac{\lambda _{n}^{2}}{\ell ^{2}}+m^{2}},
\end{equation}
 where \( \lambda  \) is the \( n \)-th real root of the transcendental
equation determined by the boundary conditions, \( \ell  \) is a
chararacteristic length along the \( {\mathcal{OX}}_{d} \) direction,
and \( m \) is the mass of an excitation of the quantum field. A
simple integral representation of \( \sum _{n}\Omega _{n} \) can
be written if we make use of Cauchy's integral formula. In fact, it
is easily seen that \begin{equation}
\label{cauchy}
\sum _{n}\Omega _{n}=-\oint _{\Gamma }\frac{dq}{2\pi }\sum _{n}\frac{2q^{2}}{q^{2}+\Omega _{n}^{2}},
\end{equation}
 where in principle \( \Gamma  \) is restricted to be a Jordan curve
on the \( q \)-complex plane with \( \Im q>0 \), consisting of a
semicircle of infinitely large radius, whose diameter is the entire
real axis. Taking (\ref{cauchy}) into (\ref{unregenergy}) we obtain
\begin{equation}
\label{unregenergy2}
E_{0}\left( d\right) =-\alpha \left( d\right) \, \frac{L^{d-1}}{2}\int \frac{d^{d-1}p_{\bot }}{(2\pi )^{d-1}}\oint _{\Gamma }\frac{dq}{2\pi }\sum _{n}\frac{2q^{2}}{q^{2}+\Omega _{n}^{2}}.
\end{equation}
 In order to proceed we must be able to perform (in a physically meaningful
way, as mentioned above) the discrete sum in (\ref{unregenergy2}).

Consider a complex function \( G(z) \) of a single complex variable
z, symmetrical on the real axis and such that its roots are simple,
non-zero and symmetrical with respect to the origin of the complex
plane. The assumption that the origin is not a root of \( G(z) \)
is not a restrictive one, because if \( z=0 \) happens to be a root
of \( G(z) \), we can always divide \( G(z) \) by some convenient
power of \( z \) in order to eliminate zero from the set of roots,
without introducing any new singularity. Let us order and count the
roots of \( G(z) \) in such a way that \begin{equation}
\label{symmetry}
\lambda _{n}=-\lambda _{-n},\; \; \; \; \; \; n=\pm 1,\pm 2,\ldots \, .
\end{equation}
 Now define the following meromorphic function \begin{equation}
J(z):=\sum _{n=-\infty }^{\infty \, \, \, \, \, \prime }\frac{1}{z-i\lambda _{n}},
\end{equation}
 where the prime indicates that the term corresponding to \( n=0 \)
must be omitted from the sum. The following properties of \( J(z) \)
are straightforward: (i) \( J(z) \) has first order poles which are
determined by the roots of \( G(iz) \): (ii) the corresponding residua
are all equal to one. Taking into account property (\ref{symmetry}),
we see that \( J(z) \) can be rewritten as \begin{eqnarray}
J(z) & = & \frac{1}{2}\left( \sum _{n=-\infty }^{\infty \, \, \, \, \, \prime }\frac{1}{z-i\lambda _{n}}+\sum _{n=-\infty }^{\infty \, \, \, \, \, \prime }\frac{1}{z+i\lambda _{n}}\right) \nonumber \\
 & = & \sum _{n=1}^{\infty }\frac{2z}{z^{2}+\lambda _{n}^{2}},
\end{eqnarray}
 Let us consider now the function \( K(z):=G(iz) \). We can state
that \begin{equation}
\label{J2}
J(z)=\frac{K^{\prime }(z)}{K(z)},
\end{equation}
 where the prime stands for `derivative with respect to \( z \)'.
In fact, the function \( K^{\prime }(z)/K(z) \) has the same simple
poles as the originally defined \( J(z) \), and also the same residuum
at each pole. Hence, we can invoke the Mittag-Leffler theorem, to
conclude that Eq. (\ref{J2}) is true up to an entire function which
does not contribute to the evaluation of the vacuum energy. It then
follows that we can write \begin{equation}
\label{sumrule}
\frac{K^{\prime }(z)}{K(z)}=\frac{d}{dz}\log \left[ K(z)\right] =\sum _{n=1}^{\infty }\frac{2z}{z^{2}+\lambda _{n}^{2}}.
\end{equation}

In order to make use of Eq. (\ref{sumrule}), we first link the dimensionless
complex variable \( z \) to the auxiliary complex momentum variable
\( q \), through the relation \begin{equation}
q^{2}+\Omega _{n}^{2}=\frac{z^{2}+\lambda _{n}^{2}}{\ell ^{2}}
\end{equation}
 hence \begin{equation}
\label{zeta}
z=z\left( q,p_{\bot }\right) =\ell \sqrt{q^{2}+p_{\bot }^{2}+m^{2}},
\end{equation}
 and \begin{equation}
\label{sumsum}
\sum _{n=1}^{\infty }\frac{2q^{2}}{q^{2}+\Omega _{n}^{2}}=\frac{\ell ^{2}q^{2}}{z}\sum _{n=1}^{\infty }\frac{2z}{z^{2}+\lambda _{n}^{2}}.
\end{equation}
 Notice that \( \lambda _{0}=0 \) is explicitly omitted on the lhs
of Eq. (\ref{sumsum}). Changing variables (\( d/dz=(z/\ell ^{2}q)d/dq \)),
we obtain for the unregularised Casimir energy the following expression
\begin{equation}
E_{0}\left( d\right) =-\alpha \left( d\right) \, \frac{L^{d-1}}{2}\int \frac{d^{d-1}p_{\bot }}{(2\pi )^{d-1}}\oint _{\Gamma }\frac{dq}{2\pi }q\frac{d}{dq}\log \left[ K(z)\right] ,
\end{equation}
 which can be integrated by parts to yield \begin{eqnarray}
E_{0}\left( d\right) &= & -\alpha \left( d\right) \, \frac{L^{d-1}}{2}\int
\frac{d^{d-1}p_{\bot }}{(2\pi )^{d-1}}\int _{\Gamma }\frac{dq}{2\pi }\frac{d}{dq}\left\{
q\log \left[ K(z)\right] \right\} \nonumber \label{unregenergy3} \\
&&+  \alpha \left( d\right) \, \frac{L^{d-1}}{2}\int
\frac{d^{d-1}p_{\bot }}{(2\pi )^{d-1}}\int _{\Gamma
}\frac{dq}{2\pi }\log \left[ K(z)\right] .\label{Eintegral}
\end{eqnarray}
 Notice that the integration is now performed on an open curve which
lies on the Riemann surface of the integrand the projection of
which on the \( q \)-complex plane is the curve \( \Gamma  \).
This curve begins at a point in the second quadrant very close to
the imaginary axis and very far from the origin, describes an arc
of a circle counterclockwise until it meets the real axis, then
proceeds to the right along the real axis towards a point very far
from the origin, where it starts to describe another arc of circle
counterclockwise until it meets again the imaginary axis (Fig. 1). 
\begin{figure}[htb]
%\vskip-3cm
\centerline{\epsfxsize=11cm \epsfbox{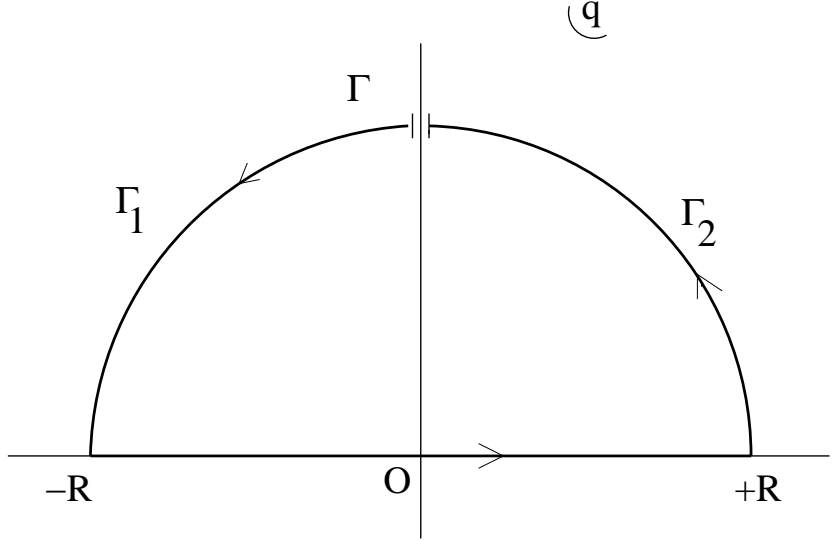}}
%\vskip-25mm
\caption{{\protect\small The curve \( \Gamma  \) begins at a point in 
the second quadrant very close to the imaginary axis and very far from 
the origin, describes an arc of a circle counterclockwise, then
proceeds to the right along the real axis towards a point very far
from the origin, where it starts to describe another arc 
 until it (almost) meets again the imaginary axis. The starting and end 
points are symmetrical with respect to the imaginary axis.
\( \Gamma _{1} \) is the part of \( \Gamma  \) that  goes from the
starting point to the origin and \( \Gamma _{2} \) is the part
from the origin to the finishing point.}} \label{f1}
\end{figure}

The
first term on the rhs of (\ref{Eintegral}) contributes with a
phase which cancels out with a phase coming from the second term,
being the final result real. The second term still needs
regularization and in order to accomplish this we split the
function \( K\left( z\right)  \) in two parts\begin{equation}
K\left( z\right) =K_{1}\left( z\right) +K_{2}\left( z\right) ,
\end{equation}
 with the following properties: (i) all terms whose integrals diverge
when \( Re\, z>0 \) are in \( K_{1} \) and (ii) all terms whose
integrals diverge when \( Re\, z<0 \) are in \( K_{2} \), (iii)
the symmetry\begin{equation}
\label{Symmetry2}
K_{1}\left( z\right) =K_{2}\left( -z\right) ,
\end{equation}
 holds. Then, we can write\begin{eqnarray}
I\,  & := & \int _{\Gamma }\frac{dq}{2\pi }\log \left[ K(z)\right] \nonumber \\
 & = & \int _{\Gamma _{1}}\frac{dq}{2\pi }\log \left[ K_{1}\left( z\right) \right] +\int _{\Gamma _{1}}\frac{dq}{2\pi }\log \left[ 1+\frac{K_{2}\left( z\right) }{K_{1}\left( z\right) }\right] \nonumber \\
 &&+  \int _{\Gamma _{2}}\frac{dq}{2\pi }\log \left[ K_{2}\left( z\right) \right] +\int _{\Gamma _{2}}\frac{dq}{2\pi }\log \left[ 1+\frac{K_{1}\left( z\right) }{K_{2}\left( z\right) }\right] ,\label{Iintegral}
\end{eqnarray}
where the integrating path has been divided into two parts: \(
\Gamma _{1} \) is the part of \( \Gamma  \) that goes from the
starting point to the origin and \( \Gamma _{2} \) is the other part,
from the origin to the end point (Fig. 1). The integrals over \( \log
K_{j}\left( z\right) ,\, j=1,2 \), are evaluated along the
different parts, \( \Gamma _{j} \), of the curve \( \Gamma  \)
and, owing to the symmetry of the construction, their sum is equal
to the integral of \( \log K_{1} \) along the total path \( \Gamma
\); this yields a phase that, as remarked above,  cancels the
phase coming from the first term in Eq. (\ref{Eintegral}). The
other two integrals are reduced to their contribution along the
real axis, and we can write
\begin{equation} \label{unregenergy4} E_{0}\left( d\right) =\alpha
\left( d\right) \, \frac{L^{d-1}}{2}\int \frac{d^{d}p}{(2\pi
)^{d}}\log \left[ 1+\frac{K_{1}\left( z\right) }{K_{2}\left(
z\right) }\right] ,
\end{equation}
 where all along the real axis \( z \) does not change its sign and
is a function of \( p:=\left( q:=p_{d},p_{\bot }\right)  \) and \( m \)
given by \( z=+\ell \sqrt{p^{2}+m^{2}} \). Notice that Eq. (\ref{unregenergy4})
gives the \emph{regularised} confined vacuum energy.

We can proceed a little bit further, still without specialising
Eq. (\ref{unregenergy4}). First we make use of
\begin{equation}
\int d^{d}p\, f\left( p\right) =\frac{2\pi ^{\frac{d}{2}}}{\Gamma
\left( \frac{d}{2}\right) }\int _{0}^{\infty }dp\, p^{d-1}f\left(
p\right)
\end{equation}
 and rescale the integration variable according to \( p\rightarrow x/\ell  \)
to obtain \begin{equation}
\label{xdenergy}
E_{0}\left( \ell ,\mu ,d\right) =\alpha \left( d\right) \frac{\, L^{d-1}}{\left( 2\pi \right) ^{d}\Gamma \left( \frac{d}{2}\right) \, \ell ^{d}}\int _{0}^{\infty }dx\, x^{d-1}\log \left[ 1+\frac{K_{1}\left( z\right) }{K_{2}\left( z\right) }\right] ,
\end{equation}
 where now \( z=+\sqrt{x^{2}+\mu ^{2}} \) with \( \mu :=m\ell  \).
Let us change the integration variable, according to \(
z\rightarrow \omega =\left( x^{2}+ \mu ^{2}\right) ^{1/2} \). Then
it follows that the regularised vacuum energy is given by
\begin{equation}
\label{mainresult}
E_{0}\left( \ell ,\mu ,d\right) =\alpha \left( d\right) \frac{\, L^{d-1}}{\left( 2\pi \right) ^{d}\Gamma \left( \frac{d}{2}\right) \, \ell ^{d}}\int _{\mu }^{\infty }d\omega \, \left( \omega ^{2}-\mu ^{2}\right) ^{\frac{d}{2}-1}\log \left[ 1+\frac{K_{1}\left( \omega \right) }{K_{2}\left( \omega \right) }\right] .
\end{equation}
Equation (\ref{mainresult}) -- or if one wishes Eq.
(\ref{unregenergy4}) -- is our main result. It is not hard to
realize that this result holds for arbitrary boundary conditions
imposed on the field on flat surfaces perpendicular to the \(
{\mathcal{OX}}_{d} \) axis. However, due to the subtraction of the
zero root from the set of all roots, an extra multiplicative
factor of 2 will be needed when dealing with topological periodic
or antiperiodic conditions. The function \( K \) can be
constructed from the boundary conditions, as we will see
explicitly in the next section.

\section{Massive fermion field under MIT boundary conditions}

As an example of the usefulness of (\ref{unregenergy4}), let us
apply it to a massive fermion field confined within an
hypothetical hyperbag with MIT boundary conditions imposed on the
field, on the \( {\mathcal{OX}}_{d} \) axis. It can be shown that,
for this relatively simple bag model, the eigenvalues of \( p_{d}
\) are determined by the roots of the function \( F(x_{d} :=
p_{d}\ell ) \), defined by \cite{MamaevTrunov80}\begin{equation}
F(p_{d}\ell )=\mu \sin (p_{d}\ell )+p_{d}\ell \cos (p_{d}\ell ),
\end{equation}
 Therefore we can choose \( G(\omega ) \) as \begin{equation}
G(\omega )=\mu \frac{\sin \left( \omega \right) }{\omega }+\cos \left( \omega \right)
\end{equation}
 where we have divided \( F(\omega ) \) by \( \omega  \) because
\( \omega =0 \) is a root of \( F\left( \omega \right)  \). We can
easily prove that for the case at hand the roots of \( G(\omega ) \)
are all real. Now, we construct \( K\left( \omega \right)  \) in
accordance with \begin{equation}
K(\omega )=G(i\omega )=\mu \frac{\sinh \left( \omega \right) }{\omega }+\cosh \left( \omega \right) .
\end{equation}
 It follows that we can write \begin{equation}
K_{1}\left( \omega \right) =\frac{1}{2}\left( 1-\frac{\mu }{\omega }\right) \, e^{-\omega },
\end{equation}
 and\begin{equation}
K_{2}\left( \omega \right) =\frac{1}{2}\left( 1+\frac{\mu }{\omega }\right) \, e^{\omega }.
\end{equation}
Notice that the symmetry given by Eq. (\ref{Symmetry2}) is obeyed
as it should. For a fermionic quantum field in \( d+1 \) dimensions,
we have \( \alpha \left( d\right) =-2\times C\left( d\right)  \),
where the factor \( 2 \) takes into account the existence of particle
and antiparticle states, while \( C\left( d\right)  \) counts the
number of different spin states and is given by \( 2^{\left( d-1\right) /2} \),
for \( d \) odd, and by \( 2^{\left( d-2\right) /2} \), for d even.
Therefore the regularised vacuum energy of the fermion field is \begin{equation}
\label{MITdenergy}
E_{0}\left( \ell ,\mu ,d\right) =-C\left( d\right) \frac{\, L^{d-1}}{2^{d-1}\pi ^{d/2}\Gamma \left( \frac{d}{2}\right) \, \ell ^{d}}\int _{\mu }^{\infty }d\omega \, \left( \omega ^{2}-\mu ^{2}\right) ^{\frac{d}{2}-1}\log \left[ 1+\frac{\omega -\mu }{\omega +\mu }\, e^{-2\omega }\right] .
\end{equation}
 Setting \( d=3 \) we readily obtain the result given in Ref. \cite{MamaevTrunov80}.
Equation (\ref{MITdenergy}) can be integrate by the following procedure.
First we make use of\begin{equation}
\log \left( 1+X\right) =-\sum _{k=1}^{\infty }\frac{\left( -X\right) ^{k}}{k},
\end{equation}
 which holds for \( -1\leq X<1 \), and write Eq. (\ref{xdenergy})
under the form \begin{equation}
E_{0}\left( \ell ,\mu ,d\right) =-\frac{C\left( d\right) \, L^{d-1}}{2^{d-1}\pi ^{d/2}\Gamma \left( \frac{d}{2}\right) \, \ell ^{d}}\sum _{k=1}^{\infty }\frac{\left( -1\right) ^{k+1}}{k}I_{k}\left( \mu ,d\right) ,
\end{equation}
 where\begin{equation}
I_{k}\left( \mu ,d\right) :=\int _{\mu }^{\infty }d\omega \, \omega \left( \omega +\mu \right) ^{\frac{d}{2}-1-k}\left( \omega -\mu \right) ^{\frac{d}{2}-1+k}e^{-2k\omega }
\end{equation}
 In order to evaluate this integral, consider the auxiliary one \begin{equation}
I_{k}\left( \mu ,d,\lambda \right) :=\int _{\mu }^{\infty }d\omega \, \left( \omega +\mu \right) ^{\frac{d}{2}-1-k}\left( \omega -\mu \right) ^{\frac{d}{2}-1+k}e^{-2k\omega \lambda },
\end{equation}
 so that\begin{equation}
I_{k}\left( \mu ,d\right) =-\frac{1}{2k}\frac{d}{d\lambda }I_{k}\left( \mu ,d,\lambda =1\right) .
\end{equation}
 To evaluate the auxiliary integral we make use of (\emph{c.f.} formula
3.384.3 in \cite{Grad94}):\begin{eqnarray}
\int _{\mu _{1}}^{\infty }dx\, \left( x+\beta \right) ^{2\nu -1}\left( x-\mu _{1}\right) ^{2\rho -1}e^{-\mu _{2}x}=\frac{\left( \mu _{1}+\beta \right) ^{\nu +\rho -1}}{\mu _{2}^{\nu +\rho }}\exp \left[ \frac{\left( \beta -\mu _{1}\right) }{2}\mu _{2}\right]  &  & \nonumber \\
\times \, \Gamma \left( 2\rho \right) W_{\nu -\rho ,\nu +\rho -\frac{1}{2}}\left( \mu _{1}\mu _{2}+\beta \mu _{2}\right) , &  &
\end{eqnarray}
 where \( W_{\nu ,\mu }\left( z\right)  \) is the Whittaker function
\cite{Whittaker}. This result holds for \( \mu _{1}>0 \), \( \left| \mbox {Arg}\left( \beta +\mu _{1}\right) \right| <\pi , \)
Re \( \mu _{2}>0 \) and Re \( \rho >0 \). The auxiliary integral
then reads \begin{equation}
I_{k}\left( \mu ,d,\lambda \right) :=-\frac{\left( 2\mu \right) ^{\frac{d}{2}-1}}{\left( 2k\lambda \right) ^{\frac{d}{2}}}\Gamma \left( \frac{d}{2}+k\right) W_{-k,\frac{d-1}{2}}\left( 4\mu k\lambda \right) .
\end{equation}
 From this result we easily obtain \begin{equation}
I_{k}\left( \mu ,d\right) =-\frac{\Gamma \left( \frac{d}{2}+k\right) \mu ^{\frac{d-2}{2}}}{4k^{\frac{d+2}{2}}}\frac{d}{d\lambda }\left[ \lambda ^{-\frac{d}{2}}\, W_{-k,\frac{d-1}{2}}\left( 4\mu k\lambda \right) \right] _{\lambda =1}.
\end{equation}
 Therefore the Casimir energy is, at last, \begin{eqnarray}
E_{0}\left( \ell ,\mu ,d\right) =-\frac{C\left( d\right) \, L^{d-1}\mu ^{\frac{d-2}{2}}}{2^{d+1}\pi ^{d/2}\Gamma \left( \frac{d}{2}\right) \, \ell ^{d}}\sum _{k=1}^{\infty }\frac{\left( -1\right) ^{k+1}}{k^{\frac{d}{2}+2}}\Gamma \left( \frac{d}{2}+k\right)  &  & \nonumber \\
\times \, \frac{d}{d\lambda }\left[ \lambda ^{-\frac{d}{2}}\, W_{-k,\frac{d-1}{2}}\left( 4\mu k\lambda \right) \right] _{\lambda =1}. &  & \label{MITenergy}
\end{eqnarray}
 Analytically, this is as far as we can go and can indeed be considered
a nice result. In order to get, however, a taste of the order of magnitude
of the different terms involved, or to get a graphical plot of the
Casimir energy, we must evaluate Eq. (\ref{MITenergy}) numerically
(which can be done very quickly) or consider some special limits.

\subsection{The limit \protect\protect\protect\protect\( \mu \ll1 \protect \protect \protect \protect \) }

The limit \( \mu \ll 1 \) is determined by the behavior of the Whittaker
function on the positive real axis for small values of the argument.
For completeness, we state some facts concerning this function and
relevant to our purposes in the Appendix. It follows then that, for
\( \mu \ll 1 \), the Casimir energy is given by \begin{equation}
\label{smallzE}
\frac{E_{0}\left( \ell ,\mu ,d\right) }{L^{d-1}}\approx -\frac{C\left( d\right) \, \left( 2^{d}-1\right) \Gamma \left( d\right) \zeta _{R}\left( d+1\right) }{2^{3d-1}\pi ^{d/2}\Gamma \left( \frac{d}{2}\right) \, \ell ^{d}}.
\end{equation}
 Setting \( d=3 \) we obtain the well known result due to Johnson
\cite{Johnson75}, namely\begin{equation}
\frac{E_{0}\left( \ell ,\mu ,d\right) }{L^{d-1}}\approx -\frac{7\pi ^{2}}{2880\, \ell ^{3}}.
\end{equation}
 Equation (\ref{smallzE}) is also in agreement with the analysis
of De Paola, Rodrigues and Svaiter for the \( d+1 \) dimensional
slab-bag for massless fermions \cite{Svaiteretal1999}. This can be
easily verified by setting \( d=D-1 \) and making use of the duplication
formula for the gamma function \cite{Grad94}\begin{equation}
\Gamma \left( 2x\right) =\frac{2^{2x-1}}{\sqrt{\pi }}\Gamma \left( x\right) \Gamma \left( x+\frac{1}{2}\right) ,
\end{equation}
 with \( x=\left( D-1\right) /2 \). For \( d=3 \), in order to compare
our result with one obtained by \cite{MamaevTrunov80} in this limit
we take into account the second term of the small \( z \) approximation
of the Whittaker function (see the Appendix). The result is then \begin{equation}
\frac{E_{0}\left( \ell ,\mu ,d\right) }{L^{d-1}}\approx -\frac{7\pi ^{2}}{2880\, \ell ^{3}}\left( 1-B\mu \right) ,
\end{equation}
 where \( B=540\, \zeta \left( 3\right) /7\pi ^{4}\approx 0.95 \)
which is approximately half of the value obtained in \cite{MamaevTrunov80}.

\subsection{The limit \protect\protect\protect\protect\( \mu \gg 1\protect \protect \protect \protect \)}

Making use of the large \( \left| z\right|  \) representation of
the Whittaker function (see the Appendix), we obtain\begin{equation}
\frac{E_{0}\left( \ell ,\mu ,d\right) }{L^{d-1}}\approx -C\left( d\right) \frac{\Gamma \left( \frac{d}{2}+1\right) }{2^{d+3}\pi ^{d/2}\Gamma \left( \frac{d}{2}\right) \ell ^{d}}\left[ \left( \frac{d}{2}+1\right) \mu ^{\frac{d}{2}-2}+2\mu ^{\frac{d}{2}-1}\right] e^{-2\mu }.
\end{equation}
 Setting \( d=3 \) and keeping only the leading term, we get \begin{equation}
\frac{E_{0}\left( \ell ,\mu ,d=3\right) }{L^{2}}\approx -\frac{3\mu ^{1/2}}{2^{5}\pi ^{3/2}\ell ^{3}}\, e^{-2\mu },
\end{equation}
 in agreement with \cite{MamaevTrunov80} as soon as we take into
account the particle and antiparticle states, and thus multiply these
authors' result by a factor of 2.

\section{Conclusions}

In this paper we have derived a general regularised expression, Eq.
(\ref{unregenergy4}), for the evaluation of the Casimir energy of
a quantum field in a flat manifold under the influence of boundary
conditions, imposed on the field on flat surfaces, or topological
conditions constraining the motion along a particular spatial direction.
Then we have applied the main result to the case of a massive fermionic
quantum field confined by a planar \( n+1 \) dimensional slab-bag
with MIT boundary conditions. An exact result for the confined fermionic
vacuum energy was obtained in terms of the derivative of the Whittaker
function with respect to a dimensionless parameter. Approximate expressions
for the vacuum energy were obtained in the limits of small and large
values of the parameter \( \mu =\ell m \).

The authors have verified explicitly that the simple Eq. (\ref{unregenergy4})
works perfectly well in several other instances. For example, it also
holds when the \( {\mathcal{OX}}_{d} \) direction is compactified
by the imposition of topological conditions, periodic or antiperiodic,
on the quantum field along that direction. In fact, the method can
be extended and applied to cylindrical and spherical geometries embedded
in \( d+1 \) dimensional spacetimes and, possibly, to other more
complex combinations of fields and geometries.

\section*{Acknowledgments}

One of the authors (A.C.T.) wishes to acknowledge the hospitality
of the Institut d'Estudis Espacials de Catalunya (IEEC/CSIC) and Universitat
de Barcelona, Departament d'Estrutura i Constituents de la Mat\`{e}ria
where this work was completed, and also the financial support of CAPES,
the Brazilian agency for faculty improvement, Grant BEX 0682/01-2.
The investigation of E.E. has been supported by DGI/SGPI (Spain),
project BFM2000-0810, and by CIRIT (Generalitat de Catalunya), contract
1999SGR-00257.

\section*{Appendix: The Whittaker function}

The Whittaker function is defined by \cite{Whittaker}\[
W_{\nu \, \sigma }\left( z\right) :=e^{-z/2}\, z^{\sigma +1/2}\, U\left( \frac{1}{2}+\sigma -\nu ,1+2\sigma ;z\right) ,\]
 where \( U\left( a,b;z\right)  \) is the confluent hypergeometric
function of the second kind, \[
U\left( a,b;z\right) =\frac{1}{\Gamma \left( a\right) }\int _{0}e^{-zt}\, t^{a-1}\left( 1+t\right) ^{b-a-1}.\]
 For the situations we are interested in, the small \( \left| z\right|  \)
behavior of this function is given by \[
U\left( a,b:z\right) \approx \frac{\Gamma \left( b-1\right) }{\Gamma \left( a\right) }z^{1-b},\, \, \, \, \, \, \, \, \, \, \, \, \, \, \, \, \, Re\, b\geq 2,\, b\neq 2.\]
 As a consequence, for small \( \left| z\right|  \), the Whittaker
function behaves as\[
W_{\nu \, \sigma }\left( z\right) \approx \frac{\Gamma \left( 2\sigma \right) }{\Gamma \left( \frac{1}{2}+\sigma -\nu \right) }\left( z^{-\sigma +1/2}-\frac{1}{2}z^{-\sigma +3/2}\right) .\]
 For large values of \( \left| z\right|  \), the behavior of the
Whittaker function is given by a series representation \cite{Grad94,Whittaker},
whose leading term is \[
W_{\nu ,\sigma }\left( z\right) \approx e^{-z/2}\, z^{\sigma }.\]

\vspace{1cm}

%\newpage


\begin{thebibliography}{10}
\bibitem{Casimir48}H. B. G. Casimir, Proc. K. Ned. Akad. Wet. \textbf{51}, (1948) 793.
\bibitem{BMohideenM2001}M. Bordag, U. Mohideen and V. M. Mostepanenko, Phys. Rept. \textbf{353}
(2001), 1. See also: G. Plunien, B. Muller and W. Greiner, Phys. Rep.
\textbf{134}, (1986) 87; V. Mostepanenko and N. N. Trunov, \emph{The
Casimir effect and its Applications}, (Clarendon Press, Oxford, 1997).
\bibitem{BagModel}A. Chodos, R. L. Jaffe, C. B. Thorn and V. Weisskopf, Phys. Rev D
\textbf{9} (1974) 3471; A. Chodos, R. L. Jaffe, C. B. Thorn, Phys.
Rev D \textbf{10} (1974), 2599; G. E. Brown and M. Rho, Phys. Lett.
B \textbf{82} (1979) 177; G. E. Brown, A. D. Jackson, M. Rho and V.
Vento, Phys. Lett. B \textbf{140} (1984) 285; K. A. Milton: Phys.
Rev. D \textbf{22} (1980) 1441; Phys. Rev. D \textbf{22} (1980) 1444;
Ann. Phys. \textbf{150} (1983) 432; J. Baacke and Y. Igarashi, Phys.
Rev. D \textbf{27} (1983); G. Cognola, E. Elizalde and K. Kirsten,
J. of Phys. A: Math. Gen. \textbf{34} (2001), 7311.
\bibitem{BagModelBC}M. De Francia, H. Falomir and E. M. Santagelo: Phys. Lett. B \textbf{371}
(1996), 285, Phys. Rev. D \textbf{45} (1992), 285. M. De Francia.
H. Falomir and M. Loewe, Phys. Rev. 55 (1997), 2477; E. Elizalde,
M. Bordag and K. Kirtsen, J. of. Phys. A: Math. Gen. \textbf{31} (1998)
1743.
\bibitem{Johnson75}K. Johnson, Acta Phys. Polonica B \textbf{6} (1975), 865.
\bibitem{Svaiteretal1999}R. D. M. De Paola, R. B. Rodrigues and N. F. Svaiter, Mod. Phys. Lett.
A \textbf{14} (1999), 2353.
\bibitem{MamaevTrunov80}S. G. Mamaev and N. N. Trunov, Izv. Vyssh. Uchebn. Zaved., Fiz. No.
7, \textbf{9} (1980). English transl. Sov. Phys. \textbf{23} (1980)
551.
\bibitem{Elisalde94}E. Elizalde, S. D. Odintsov, A. Romeo, A. A. Bytsenko and S. Zerbini,
\textit{Zeta function regularization techniques with applications},
(World Scientific, Singapore, 1994). E. Elizalde: \emph{Ten Physical
Applications of Spectral Zeta Functions}, (Springer-Verlag, Berlin,
1995).
\bibitem{relatedpapers}M. Bordag, E. Elizalde and K. Kirsten, J. of Math. Phys. \textbf{37},
(1996) 895: M. Bordag, E. Elizalde, B. Geyer and K. Kirsten, Comm.
Math. Phys. \textbf{179}, (1996) 215; M. Bordag, E. Elizalde, K. Kirsten
and S. Leseduarte, Phys. Rev. D \textbf{56}, (1997) 4896; M. Bordag,
E. Elizalde and K. Kirsten, J. of Phys. A: Math. Gen. \textbf{31},
(1998) 1743.
\bibitem{ST2002}F. C. Santos and A. C. Tort, \emph{The Casimir energy of a massive
fermion field revisited}, quant-ph/0201104.
\bibitem{FO2001}K. Fukushima and K. Ohta, Physica A \textbf{299} (2001) 48.
\bibitem{Grad94}I. S. Gradshteyn and I. M. Ryzhik, \emph{Table of Integrals, Series,
and Products}, 5th ed. (Academic Press, San Diego, 1994).
\bibitem{Whittaker}E. T. Whittaker and G. N. Watson, \emph{A Course in Modern Analysis,}
4th ed. (Cambridge University Press, Cambridge, 1990 ); M. Abramowitz
and I. Stegun, \emph{Handbook of Mathematical Functions with Formulas,
Graphs and Mathematical Tables}, 9th printing, (Dover, New York, 1972). \end{thebibliography}
\end{document}